\documentclass[sigconf]{acmart}
\AtBeginDocument{%
  }

\copyrightyear{2025}
\acmYear{2025}
\setcopyright{acmlicensed}
\acmConference[SIGIR '25]{Proceedings of the 48th International ACM SIGIR Conference on Research and Development in Information Retrieval}{July 13--18, 2025}{Padua, Italy}
\acmBooktitle{Proceedings of the 48th International ACM SIGIR Conference on Research and Development in Information Retrieval (SIGIR '25), July 13--18, 2025, Padua, Italy}
\acmDOI{10.1145/3726302.3731962}
\acmISBN{979-8-4007-1592-1/2025/07}  

\usepackage[utf8]{inputenc}
\renewcommand\footnotetextcopyrightpermission[1]{} 

\usepackage{graphicx}

\usepackage[linesnumbered,ruled,vlined]{algorithm2e}
 
\usepackage[utf8]{inputenc}

\usepackage{amssymb}
\usepackage{xcolor}
\usepackage{pifont} 
\usepackage[T1]{fontenc}

\begin{document}

\title{Content Moderation in TV Search: Balancing Policy Compliance, Relevance, and User Experience}

\author{Adeep Hande}
\affiliation{
\institution{Applied AI Research, Comcast}
\city{Washington, DC}
\country{USA}}
\email{adeep\_hande2@comcast.com} 
\author{Kishorekumar Sundararajan}
\affiliation{
\institution{Applied AI Research, Comcast}
\city{Washington, DC}
\country{USA}}
\email{kishore\_kumar@comcast.com}

\author{Sardar Hamidian}
\affiliation{
\institution{Applied AI Research, Comcast}
\city{Washington, DC}
\country{USA}
}
\email{sardar\_hamidian@comcast.com}
\author{Ferhan Ture}
\affiliation{%
  \institution{Comcast AI Technologies}
  \city{Washington, DC}
  \country{USA}}
\email{ferhan\_ture@comcast.com}
\renewcommand{\shortauthors}{Adeep Hande, Kishorekumar Sundararajan, Sardar Hamidian, and Ferhan Ture}
\begin{abstract}

Millions of people rely on search functionality to find and explore content on entertainment platforms. Modern search systems use a combination of candidate generation and ranking approaches, with advanced methods leveraging deep learning and LLM-based techniques to retrieve, generate, and categorize search results. Despite these advancements, search algorithms can still surface inappropriate or irrelevant content due to factors like model unpredictability, metadata errors, or overlooked design flaws. Such issues can misalign with product goals and user expectations, potentially harming user trust and business outcomes. In this work, we introduce an additional monitoring layer using Large Language Models (LLMs) to enhance content moderation. This additional layer flags content if the user did not intend to search for it. This approach serves as a baseline for product quality assurance, with collected feedback used to refine the initial retrieval mechanisms of the search model, ensuring a safer and more reliable user experience. 
\end{abstract} 





\maketitle   
\section{Introduction}

The rapid expansion of digital TV content and the increasing adoption of text and voice-activated searches on entertainment platforms have amplified the challenges of content moderation \cite{allam2021future}. Unlike traditional systems where content that is deemed to be problematic may be outright removed, entertainment platforms have to retain the complete catalog to harbor diverse user preference across the whole customer-base --even when some items contain sensitive or potentially offensive material. This issue is particularly pertinent to TV search environments, where the user has access to a broad range of programming options on TV such as TV shows, Sports, and movies. While existing filtering mechanisms such as age-based restrictions or parental control settings operate by predefined rules that block certain categories of content, our approach explores beyond static filtering by considering contextual appropriateness based on user queries and retrieved results, to flag any entity that may be deemed sexist, offensive or xenophobic concerning the user intent.

Consequently, the challenge lies not in removing content altogether but in developing a system that can proactively filter search results, without altering the parent catalog, and allowing only contextually appropriate content in the search results. This delicate balance is essential to enhancing user satisfaction by retrieving the best search results relevant to the users' queries while adhering to the content policies and guidelines \cite{yum2024influence}. In addition to the inherent challenges of context-aware filtering, the task is further exacerbated by faulty or inconsistent metadata, which can potentially lead to misclassified search results that appear relevant but are in fact, inappropriate. Furthermore, the evolving semantics of the language in user queries \cite{karjus2024evolving}. Words and phrases that were once considered benign and appropriate may have newer connotations over time, depending on cultural and social shifts \cite{rodman2020timely, jeresano2022digital}. This dynamic nature of language evolution means that having a static set of filtering rules becomes obsolete over time, underscoring the need for an adaptive system.  It is worth noting that, although most existing search and retrieval approaches aim to mitigate these anomalies during model training \cite{chen2023bias} and data preprocessing, certain cases remain unaddressed. This additional degree of assurance is designed to capture unavoidable experiences that the initial model failed to account for.

\begin{table*}[ht]
\centering
\resizebox{\textwidth}{!}{%
\begin{tabular}{|l|p{6cm}|c|p{6cm}|}
\hline
\textbf{Search Command}             & \textbf{Search Result}                              & \textbf{Label} & \textbf{Reason}                                                                               \\ \hline
Church songs               & Documentary on church scandals               & \textcolor{red}{\ding{55}}           & Mismatch with user intent; highlights controversies instead of requested content.            \\  
Women empowerment          & Movies depicting toxic female stereotypes or sexual orientation    & \textcolor{red}{\ding{55}}           & Contradicts the theme of the user query; reinforces negative stereotypes.               \\  
Children's shows        & TV series with adult themes            & \textcolor{red}{\ding{55}}           & Inappropriate for the intended audience; including content unsuitable for children.            \\ 
Educational shows          & Science documentaries for students           & \textcolor{green}{\ding{51}}         & Aligns with the user intent; provides age-appropriate and relevant educational content.        \\ \hline
\end{tabular}%
}
\caption{Examples of flagged results with corresponding flags and reasons. Note: The queries and results displayed here are representative of the thematic content of the flagged results and not the actual text, which is omitted to maintain user privacy and avoid potential inaccuracies or legal concerns. \textcolor{red}{\ding{55}} indicates a flagged result, while \textcolor{green}{\ding{51}} denotes acceptable results.}
\label{tab:anom_results}
\end{table*}
 
Our framework operates in a dual-stage process. The primary stage applies structured meta-heuristic filtering, as shown in Algorithm \ref{alg1}, leveraging predefined lexicons and dynamically updated sensitivity scores to flag potentially inappropriate search results - ensuring the content remains accessible while aligning with platform policies. The secondary stage incorporates a large language model (LLM) \cite{grattafiori2024llama3herdmodels} to validate and refine real-time flagged results, ensuring adaptability to evolving user intent and language semantics.  To clarify filtering decisions, sensitivity scores dynamically adjust based on real-time flagged instances, integrating historical patterns and validation feedback. In contrast to static rules, this approach continuously refines lexicons and scoring mechanisms, rather than generating entirely new lexicons. By integrating LLM validation, this framework enhances precision and accuracy, ensuring that content moderation remains scalable and policy-compliant while retaining a complete catalog of user requests.

As content appropriateness is heavily influenced by user intent and context, no benchmark datasets or directly comparable models exist for evaluation. To ensure consistent and policy-compliant moderation, flagged search results undergo offline evaluation, where previous-day queries are evaluated, and all flagged instances are reviewed by an editorial team before any final action is taken. While this remains a proactive approach, the continuous feedback loop from editorial review further refines filtering over time. This framework ensures content retrieved remains aligned with platform guidelines while preserving user experience. This iterative lexicon refinement also serves as a foundation for transitioning to a smaller, distilled model, ultimately enabling real-time content moderation without on-demand LLM inference.

\begin{figure*}
  \centering
  \includegraphics[width=\linewidth]{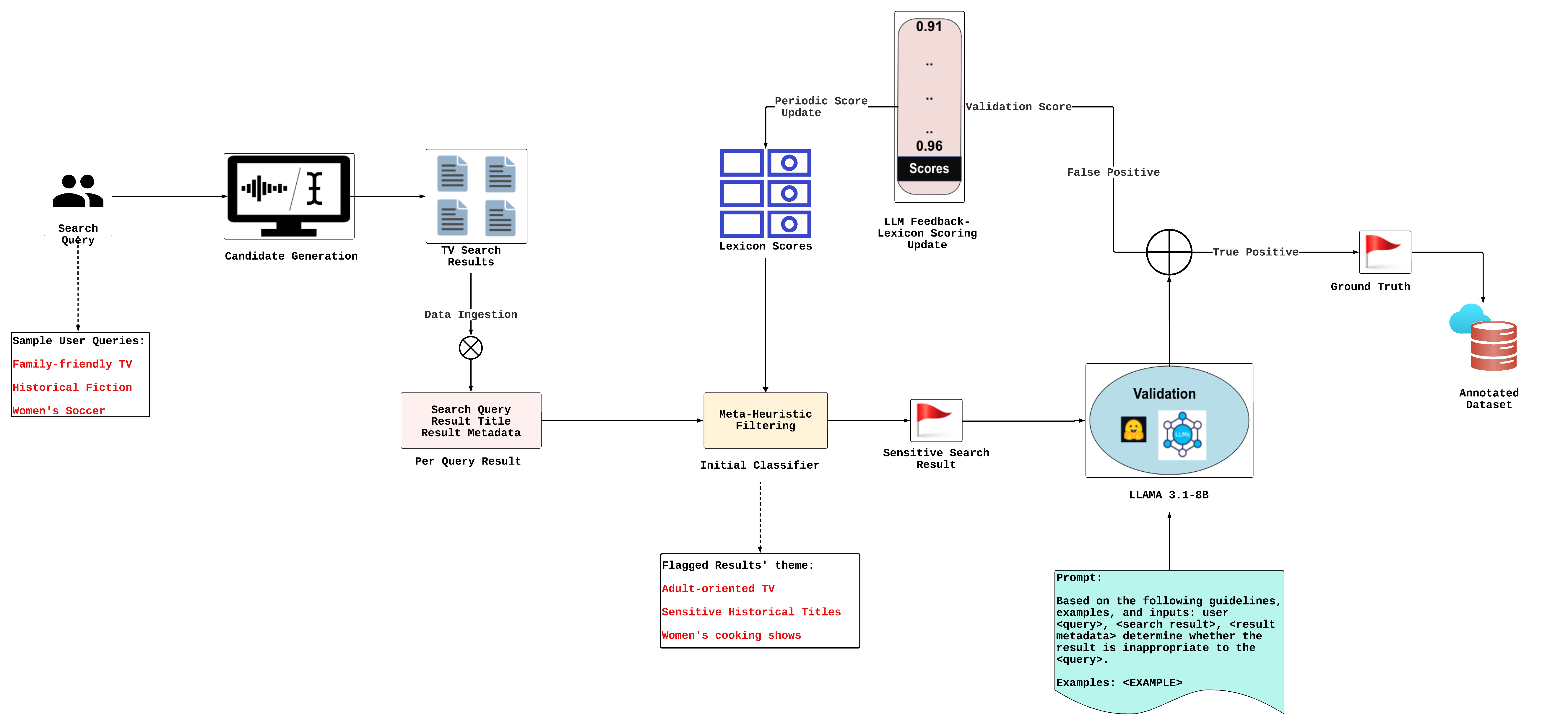}
  \caption{System overview of the filtering pipeline. Queries and results undergo metadata extraction, heuristic filtering, and LLM validation. True positives refine moderation, while false positives update sensitivity scores. Flagged results are thematic representations; titles omitted for privacy. LLM: LLAMA-3.1-8B \cite{grattafiori2024llama3herdmodels}}
  \label{fig:overview}
\end{figure*}
 
\section{Methodology}
\label{sec:methodology}
\begin{algorithm}
\caption{Meta Heuristic Filtering (Temporal Adaptation)}
\label{alg1}
\KwIn{Query $Q$, Results $R$, Metadata $M$, Lexicons $L$, Time $t$, \\
\quad Similarity Threshold $T_s$, Flagging Threshold $\beta$}
\KwOut{Flagged Set $F$}

\SetKwFunction{ComputeScore}{g} 

\SetKwProg{Fn}{Function}{:}{}
\Fn{\ComputeScore{$x$}}{
    $S_{\text{total}} \gets 0$\;
    $N(x) \gets |x|$\;
    \ForEach{$L_j \in x$}{
    \Return $\dfrac{\sum_{L_j \in x} S(L_j, t)}{N(x)}$ \; \tcp{ Computes the average sensitivity score for query, result, or metadata}
}}

\Fn{$S(L_j, t)$}{ 
    \Return $\alpha \Big(1 - \dfrac{f(L_j,0)}{\sum_k f(L_k,0)}\Big) + (1-\alpha)\dfrac{f(L_j,t)}{\sum_k f(L_k,t)}$\; \tcp{(2) Time-adaptive lexicon scoring function}
}

$F \gets \emptyset$\;

\ForEach{$r \in R$}{
    Compute embedding $v_Q \gets \text{Embed}(Q)$\;
    Compute embedding $v_r \gets \text{Embed}(r, M)$\;
    $s \gets \cos(v_Q, v_r)$\;
    
    \If{$s \geq T_s$}{
        $g(Q) \gets \ComputeScore(Q)$\; \tcp{Compute query score after cosine check}
        
        $g(r) \gets \ComputeScore(r)$ \tcp{Compute score for result}
        $g(m) \gets \ComputeScore(m)$ \tcp{Compute score for metadata}

        \If{$g(Q) < \beta$ \textbf{and} ($g(r) > \beta$ \textbf{or} $g(m) > \beta$)}{
            $F \gets F \cup \{r\}$\;
        }
    }
}
\Return $F$\;

\end{algorithm}

Our framework moderates TV searches by selectively flagging content that may be relevant but contextually inappropriate. Traditional search candidate generation algorithms prioritize retrieval based on embedding-based or lexical similarity, often overlooking subtle misalignments between query intent and search results. Our approach serves as an additional layer over the search algorithms to ensure more effective filtering. Building upon the existing research in transformer-based embeddings, the methodology amalgamates these components into the system in a cohesive framework. The system utilizes pretrained transformer-based embeddings to encode search queries (\textbf{Q}), results (\textbf{R}), and associated Metadata (\textbf{M}) into dense semantic representations \cite{devlin2018bert, vaswani2017attention}. Here, \textbf{M} refers to the additional information accompanying the search result - such as age rating, program genre, or content descriptions - to further help contextualize and analyze the result.  At the token level, these embeddings capture contextual relationships, enabling nuanced understanding of user intent and result relevance. We compute cosine similarity between these embeddings to filter our exact or redundant matches, thus focusing on search results where there might be a misalignment.

\subsection{Meta-Heuristic Filtering} 
Our approach leverages a set of predefined lexicons -- a curated list of user-annotated words or phrases identified as sensitive (terms that may reference racist, sexist or offensive connotations). These lexicons are organized based on domain expert input and user annotations and are continuously updated. The system then calculates \textbf{sensitivity scores}. In our framework, a \textbf{flagged instance} is defined as a search result while potentially relevant, may be contextually inappropriate concerning the users' intent, as shown in Table \ref{tab:anom_results}. 

Algorithm \ref{alg1} explains our meta-heuristic filtering approach, which dynamically evaluates query-result pairs for content moderation. Initially, the system computes embeddings for queries and results, applying a predefined similarity threshold to eliminate highly similar matches and focus on potential flags. Sensitivity scores for each lexicon are then computed based on their real-time occurrences, adjusting dynamically based on how increasingly or decreasingly it is used by customers in their search queries. These scores are then aggregated within each query, result, and metadata to evaluate the context appropriateness. A result is flagged as sensitive if its sensitivity score exceeds a predefined threshold while the corresponding query remains below it. By incorporating time as an additional modality, our framework ensures robust adaptation to language evolution. While these steps could be integrated into the retrieval process, an independent moderation layer ensures consistency despite evolving candidate generation algorithms' updates. Separating moderation from the retrieval process stabilizes policy enforcement, preventing inconsistencies in filtering outcomes. Additionally, to avoid over-filtering and overreliance on meta-heuristics, we employ LLMs for evaluation and feedback.

\subsection{LLMs as Validators} Large language models have been increasingly leveraged as validation mechanisms across various NLP applications, including knowledge graphs \cite{boylan2024kgvalidator}, software systems \cite{chen2024chatunitest, eghbali2024hallucinator}, recommender systems \cite{xu2024openp5}. These studies leverage LLMs' contextual understanding and reasoning capabilities to validate system-generated outputs. Unlike static classifiers, LLMs assign validation confidence scores to flagged instances while simultaneously performing auxiliary tasks that improve contextual reasoning, such as determining query irrelevancy, estimating users' age based on the query, detecting policy violations, and leveraging chain-of-thoughts (CoT) prompting \cite{wei2022chain}. Based on the query-result-metadata triplet, the LLM computes a validation score V as shown in Equation \ref{eq7}, where $w_i$ represents the weights for the respective tasks:

\begin{equation}
\label{eq7}
    V(Q, R_i) = \sum_{i=1}^{p} w_{p} \cdot LLM_{p}(Q, R_i, M_i) 
\end{equation}

The LLM feedback mechanism refines the lexicon sensitivity scores by incorporating validation scores from potential false positives. Each lexicon's score is updated by appending a batch-averaged validation feedback. The updated sensitivity/impact relevance equation, based on equation \ref{eq2} is:
\begin{equation}
\label{eq2}
    S(L_i, t+1) = \alpha \cdot S(L_i, t) + (1-\alpha) \cdot (1- \frac{1}{|B|}V(L_i))
\end{equation}

$\alpha$ controls historical score retention and $V(L_i)$ represents the average validation score for all query-result pairs that contain the lexicon $L_i$. Here, $|B|$ denotes the batch size,  ensuring that validation updates are stabilized over multiple samples, mitigating noise from individual false positives. This feedback loop dynamically refines filtering, ensuring adaptive lexicon scoring and robust content moderation. The system incrementally constructs an annotated dataset by processing through real-time user queries with classifier and LLM validation (LLAMA, \cite{touvron2023llama}) as shown in Fig \ref{fig:overview}.



\section{Experiments and Results}
At initialization (\textit{t=0}), the system starts with no pre-existing labeled data, relying solely on a minimal set of user-reported search results. As the system operates, search results flagged by meta-heuristic filtering are validated by the LLM, gradually expanding the dataset, gradually expanding the dataset, as shown in Table \ref{tab:anom_results}. The system prioritizes  \textit{top-k} (k=5) retrieved search results, emphasizing those most visible to users. As flagged results accumulate, the LLM feedback dynamically refines sensitivity scores and adjusts weights based on the query occurrences. If a lexicon is frequently validated as a false positive, its sensitivity score decreases, making it less likely to be detected in future queries. Filtering thresholds are continually adjusted based on the distribution of flagged instances, ensuring precision improves without over-filtering. This iterative refinement system transforms it from a reactive framework - dependent on user reporting - to a proactive framework that learns from evolving user search behaviors.  
\textbf{Parameter Selection and Optimization}: To ensure stable filtering and adaptive content moderation, key parameters $\alpha$, $\beta$, and batch size $|B|$ - were tuned empirically based on offline system performance data. The weighting factor $\alpha$ balances historical lexicon scores' relevancy with real-time LLM evaluation feedback, preventing sudden shifts in sensitivity scoring. The filtering threshold $\beta$, which determines the sensitivity score cutoff in the meta heuristic in Algorithm \ref{alg1}, was set empirically with human intervention, where domain experts reviewed flagged instances to establish a reasonable boundary between acceptable and potentially sensitive results. The parameter selection process was data-driven, relying on offline validation data from past system runs, where performance was assessed using previously logged search queries and flagged results. The batch size $|B|$ is dynamically adjusted based on the flagged results in the initial classifier, stabilizing lexicon updates over multiple iterations. These parameters were continually optimized through empirical evaluations, ensuring adaptability and preventing over-fitting. We use LLAMA 3.1-7B for evaluations \cite{inan2023llama}.\\
\textbf{Automating Content Moderation in TV search:} Traditional content moderation in this context relied on manual user reports, leading to limited detection coverage and delayed response times. Our system automates this process, identifying and refining flagged results at scale, handling close to a million search queries daily, while flagging and refining results to ensure policy compliance. At initialization, the system flagged 2,948 results, significantly increasing detection compared to manual reporting. Over eight weeks, validated flagged results increased from 168 to 371, as seen in Table \ref{tab:anom_results}. Over an eight-week evaluation, the system collected 1,814 \textit{query-result} pairs validated as true positives (TPs) by LLM evaluation. In contrast, prior to this system, content moderation relied solely on user reports, which identified only a handful of problematic cases over the same timeframe.
\begin{table}[!h]
    \centering
\begin{tabular}{|c|c|c|c|c|c|}
\hline
\textbf{Week} & \textbf{ Anomalies} & \textbf{TP} & \textbf{FP} & \textbf{Precision} & \textbf{F1} \\ \hline
1 & 2,948 & 168 & 2,780 & 0.06  & 0.11 \\  
2 & 3,154 & 230 & 2,924 & 0.07 &  0.14 \\   
3 & 4,148 & 171 & 3,977 & 0.04 & 0.08 \\  
4 & 3,603 & 211 & 3,392 & 0.06 &0.11 \\  
5 & 3,546 & 216 & 3,330 & 0.06 &  0.11 \\  
6 & 3,135 & 234 & 2,901 & 0.07 & 0.14 \\ 
7 & 2,958 & 213 & 2,745 & 0.07 & 0.13 \\  
8 & 2,510 & \textbf{371} & 2,139 & 0.15   &  0.26  \\ \hline
    \end{tabular}
    \caption{Relative Performance Metrics for the Hybrid System. TP = True Positives, FP = False Positives}
    \label{tab:weekly_metrics}
\end{table}   
\textbf{Baseline for Context-Aware Filtering:} Unlike age-based filtering systems that enforce static rules, this system establishes a baseline for context-aware filtering in TV search. No existing benchmark datasets or directly comparable models address this challenge, as filtering must adapt to the user intent, content policies and catalog changes. By integrating heuristic filtering with LLM validation and feedback, this approach offers a scalable framework for real-time content moderation. The system processes nearly a million search queries daily on average, continuously refining its filtering precision. Initially, over-weighted lexicons led to excessive flagging, but LLM validation dynamically adjusted sensitivity scores, reducing false positives. Beyond automated moderation, the system generates high-quality labeled data, supporting the continuous improvement of user experience, lexicon updates, and filtering. By refining lexicons iteratively, the system progressively reduces its dependency on the LLM for validation, paving the way for a smaller, distilled model that can replicate the evaluation process in an online setting. This transition would make the framework even more proactive without requiring on-demand LLM inference, making the system more scalable and efficient for large-scale deployment.

\section{Conclusion}
This work establishes a baseline for retrieval-aware content moderation in TV searches, specifically addressing cases where search results, while relevant, may be deemed inappropriate in the context of user intent. Unlike traditional content moderation, which relies on predefined filters, our system dynamically evaluates search result relevance and sensitivity using meta-heuristics and LLM validation. By integrating heuristics with real-time feedback, the system refines content alignment without over-filtering intended results. Over an eight-week evaluation, it demonstrated progressive improvements in filtering precision, ensuring compliance while prioritizing user experience. Since no benchmarks exist for this task, our work serves as a foundation for future content moderation in TV search. 

\section{Presenter Bio}
\textbf{Kishorekumar Sundararajan} is a Director of Software Engineering at Comcast, leading quality engineering platforms across voice, content discovery, personalization, and site reliability engineering (SRE). His work focuses on evaluating ML systems for NLP and speech, with an emphasis on relevance, quality, and responsible AI. Outside work, he enjoys road trips with his family and exploring nature.
\\
\textbf{Adeep Hande} is a Machine Learning Engineer at Comcast, specializing in quality assurance and responsible AI. He currently works on the QA team, implementing and evaluating AI and ML solutions across voice and content discovery. Adeep graduated with an MS in Data Science from Indiana University Bloomington in 2024. Outside of work, he enjoys Formula 1, traveling, hiking, and reading science fiction.

\bibliographystyle{ACM-Reference-Format}
\bibliography{sample-base,custom}


\end{document}